# Nano-jet Related to Bessel Beams and to Super-resolutions in Microsphere Optical Experiments

Yacob Ben-Aryeh·

·Correspondence

phr65yb@physics.technion.ac.il

*Physics Department,*

*Technion-Israel Institute of Technology,*

*Haifa 32000, Israel*

**Abstract**

The appearance of a Nano-jet in the microsphere optical experiments is analyzed by relating this effect to non-diffracting Bessel beams. By inserting a circular aperture with a radius of subwavelength dimension in the EM waist, and sending the transmitted light into a confocal microscope, the EM oscillations by the different Bessel beams are avoided. On this constant EM field evanescent waves are superposed. While this effect improves the optical-depth of the imaging process, the object fine-structures are obtained, from the modulation of the EM fields by the evanescent waves. The use of a combination of the microsphere optical system with an interferometer for phase contrast measurements is described.

**Keywords:** Microsphere, Nano-jet, Bessel beams, Super-resolution, Evanescent Waves, Phase Contrast

## Introduction

Researchers spent much effort in pushing the frontier of optical resolution in a quest to image very small objects. Unfortunate optical resolution is limited by diffraction and dispersion effects. Diffraction causes light beams to spread in transverse direction during their propagation. A temporal pulse of light analogously spreads owing to material dispersion. The quest for ultimate super resolution has been described in a short Review [1], treating incoherent super-resolution techniques, coherent super resolutions schemes, near field super-resolution techniques, etc. Quite long ago Silberberg [2] has shown that it is possible for self-focusing nonlinearity to compensate both spatial and temporal spreading of the light pulse. The resulting wave packet is a spatiotemporal soliton which is referred to as a "light bullet"). Minardy et al. [3], by using arrays of wave guides, demonstrated a nonlinear propagation of a 3D localized optical wave packet. Belic et al. [4] obtained exact spatiotemporal periodic travelling wave solutions to the generalized (3+1)–dimensional nonlinear Schrodinger equation and used these solutions to construct analytical light bullet solutions. Scattering effects have been found to be an impediment to focusing and imaging. Light scattering, however, does not lead to an irretrievable loss of information. The information might be scrambled into disordered interference patterns called laser speckles. Para-axial speckle based metrology systems have been analyzed by Kelly et al. [5]. There are various techniques for imaging inside scattering media. Methods for controlling waves in space and time for imaging and focusing in complex media have been developed by Mosk et al. [6]. A different approach to super resolution phenomena has been discussed in relation to microsphere optical experiments (see e.g. [7-10]). The microsphere optical system is different from those described in [1-6], as in this system the information on the fine structures of the object is obtained from the evanescent waves transmitted through the microsphere. Also as the microsphere is composed of dielectric materials there are not any nonlinear effects in this system. High resolutions were obtained in microsphere optical experiments (up to 50 nm in [7-9] and 25 nm in [10]). Such high resolutions might have an impact in the fields of integrated optics, microchips, photo lithography plus plasma etching, photo resists etc., in which very high resolutions are needed. Therefore the improvement of resolutions in these systems is of utmost importance.

There are various different studies about the high resolutions obtained in microsphere optical experiments. Pang et al. [11] and Sundaram et al. [12] claimed that the physical mechanism by which high resolutions are obtained in microsphere experiments is unclear. Hao et al. [13] described the microsphere as a channel that connects the near field evanescent waves and the transmission one in the far field. In some works by the present author [14-17] the enhancement of resolutions which are higher than the Abbe limit has been related to the field of scanning near-field optical microscopy (SNOM) [18]. In this field evanescent waves are produced in which one component of the optical wave vector is imaginary (e.g. in the symmetric $z$ axis), leading to a decay of the wave in this direction. Other components of the wave vector increase according to the Helmholtz equation, thus decreasing the effective wavelength and correspondingly increase the resolution. In previous work [17] an estimate to the increase of resolution relative to the Abbe limit has been given, and the condition for converting the evanescent to propagating waves has been derived. It has been shown in various works [19-23] that the increase of the refractive index of microspheres can enhance the imaging resolution and quality. The quantitative analysis made in previous work [17] is in a good agreement with these results.

There are special effects which were obtained with microspheres. It has been shown that by using microsphere-chain waveguides [24, 25] focusing and resolutions can improved. Microsphere near-field nano-structures were observed and analyzed using picosecond pulses [26]. Viruses were observed by microsphere optical nanoscopy [27]. It has been shown that it is possible to control the focusing properties of the microsphere by using pupil masks [28]. Resolution phase contrast imaging by digital holography has been developed and observed [29]. There are various articles relating the microsphere high resolutions to Nano-jets where optical beams are produced near the focal point which have a very narrow waist [30-34]. Appearance of Nano-jets in the microsphere optical experiments can be analyzed by the use of the Mie theory which shows certain deviations from the simple geometric optics approach [31]. This method does not give analytical results, as it requires the summation of a large number of terms for moderate sphere size. An analytical description of the microsphere axially symmetric focusing with strong aberration has been developed by the use of the Bessoid integral [35] and the results are in agreement with calculations by Mie theory. However, such calculations do not explain the high resolutions obtained in microsphere optical experiments.

The main issue of the present paper is show how to avoid diffraction effects in microsphere optical experiments. We describe a combination of a microsphere with an orifice of subwavelength dimensions in the focal plane to affect the profile of the out coming classical electromagnetic (EM) wave-front which will improve the resolution. (Such an orifice was described and constructed in [36]). While geometric optics describes the mean propagation of light the diffraction effects are described as spherical aberration. The full description of the diffraction effects in this system is very complicated [35].

However, by introducing a subwavelength aperture in the focal plane the imaging process is described by the use of Bessel function of zero order [37-39]. While usually a Bessel function is quite broad by restricting the transmitted EM beam to the orifice, the diffraction effects are eliminated and a very good optical depth is obtained. By using the present approach we analyze also the use of a combination of the microsphere with an interferometer for phase contrast measurements.

**Microsphere optical Nano-jet related to non-diffractive Bessel beams**

Production of the Nano-jet in microsphere experiments is analyzed by superposing the spherical aberration on the geometric optics trajectories, as described in Fig. 1. This Figure describes a microsphere with refractive index $n_2$ and radius R located on a thin film of object at a contact point $O$. Plane EM waves are transmitted through the object and converging by the microsphere into the focal plane. The EM field is composed of two parts: a) Evanescent waves which include the fine structures of the object (e.g. corrugated metallic film [15]) and which are decaying along the $z$ axis. These waves are transformed to propagating waves by the microsphere, [14-17] but this process is effective only near the contact point $O$ where the distance between the object plane surface and the microsphere surface is enough small so that significant amount of evanescent waves arrives at the microsphere surface. b) There is a large amount of EM waves which are not converted by the object into evanescent waves and by converging near the focal point $P$ are producing the Nano-jet. These EM waves do not include the information on the fine structure of the object but produce a constant background light intensity on the EM waist which is modulated by the evanescent waves. We describe the axially spherical symmetric focusing by superposing the Bessel functions solutions on the geometric optics trajectories. The geometric optics trajectories of the propagating waves (i.e. those which are not converted by the object into evanescent waves) are obtained as function of the incident angle $\theta_{in}$.

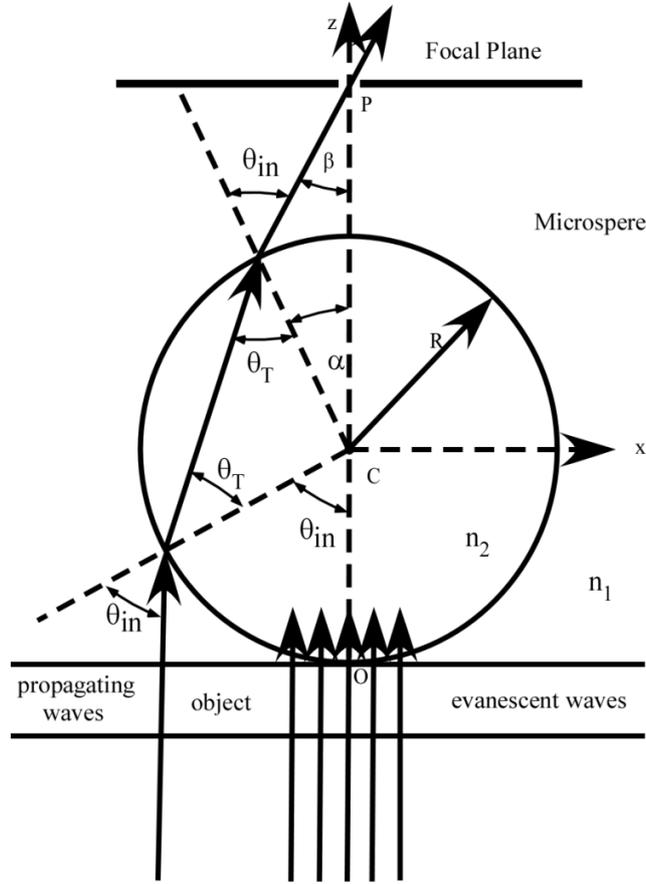

**Fig. 1**

Plane EM waves are transmitted through the object, converging by the microsphere and exiting through a small circular aperture at point $P$ located on a plane barrier at the focal plane. Evanescent waves which include the information on the fine structures of the object are transmitted into the microsphere near the contact point $O$ and are transformed by the microsphere into propagating waves. The geometric optics trajectory for a propagating ray, which is not evanescent, and which enters the microsphere far from the contact point $O$ is described with the locations of all relevant angles: $\theta_{in}$, $\theta_T$, $\alpha$ and $\beta$. Bessel beams producing the Nano-jet are obtained by the superposition of the spherical aberration on the geometric optics trajectory. The Nano-jet is modulated by the evanescent waves and the fine structures of the object are obtained by sending the radiation transmitted around point P into a confocal microscope.

According to Snell's law the incidence angle $\theta_{in}$ and the transmitted angle $\theta_T$ are related by: $n_1 \sin\theta_{in} = n_2 \sin\theta_T$, where $n_1$ is the refractive index of the medium in which the microsphere is inserted. As shown in Fig. 1 the propagating optical ray is transmitted through the first microsphere surface and afterwards transmitted on the other side of the microsphere following inverted Snell's law: The incident angle is $\theta_T$ and the transmitted angle is $\theta_{in}$. The optical ray transmitted through the microsphere is incident on the focal plane and intersects the symmetric $z$ axis with crossing angle $\beta$. As shown in Fig. 1 we have the relations $\theta_{in} - \alpha = \beta$ ; $\theta_{in} + \alpha = 2\theta_T$. Thus, we obtain

$$\beta = 2\theta_{in} - 2\theta_T \tag{1}$$

Due to the spherical symmetry of the microsphere the propagating EM waves with incident angles between $\theta_{in}$ and $\theta_{in} + d\theta_{in}$ produce on the microsphere surface a narrow spherical envelope with an area

$$dS_{sphere} = 2\pi R^2 \sin\theta_{in} d\theta_{in} \tag{2}$$

From the area $dS_{sphere}$ the EM waves are transmitted through the microsphere and are incident on the focal plane with intersecting angle $\beta$ given by eq. 1. These EM waves, which are propagating from the area $dS_{sphere}$, produce on the focal plane field intensity given by

$$\Delta E = C 2\pi R^2 \sin(\theta_{in}) \cos(\theta_{in}) t(\theta_{in}) d\theta_{in} \tag{3}$$

Here the additional trigonometric function $\cos\theta_{in}$ represents the decrease of the field intensity per unit area in the plane tangential to the microsphere surface relative to the field intensity per unit area in the $x, y$ plane. C is proportionality constant and $t(\theta_{in})$ represents the multiplication of the transmission coefficient of the object by the transmission coefficient of the microsphere. The EM field which is transmitted through the microsphere as non-evanescent wave for the angles between $\theta_{in}$ and $\theta_{in} + d\theta_{in}$ can be written in a short notation as

$$\Delta E = F(\theta_{in}) d\theta_{in} \quad ; \quad F(\theta_{in}) = C\pi R^2 \sin(2\theta_{in}) t(\theta_{in}) d\theta_{in} \tag{4}$$

Here $F(\theta_{in})$ obtains its maximal value at $\theta_{in} = 45^0$. The EM field $\Delta E$ is transmitted through a circular aperture at point P located on a plane barrier at the focal plane and it is intersecting with the symmetric z axis with an angle $\beta$, which is given by eq. 1 and Snell's law as

$$\beta = 2\theta_{in} - 2\arcsin(n_1 \sin\theta_{in}/n_2) \quad ; \quad \arcsin(n_1 \sin\theta_{in}/n_2) \equiv \theta_T \tag{5}$$

So far, we have used a geometric optics description. The present idea is that we have to take into account the spherical aberration by superposing it on the geometric optics trajectory. Then, eq. 4 is changed to

$$\Delta E = F(\theta_{in})d\theta_{in}\psi(\rho,z,t) \tag{6}$$

Here $\psi(\rho,z,t)$ is the physical wave function in cylindrical coordinates representing the spherical aberration superposed on the geometric optics differential intensity $F(\theta_{in})d\theta_{in}$. As the spherical aberration depends on the angle $\beta$ we need to solve the physical optics wave function $\psi(\rho,z,t)$ as function of $\beta$ for each $\Delta E$ of eq. 4. The problem is to deduce the form of $\psi(\rho,z,t)$ and all relevant parameters. We relate the scalar wave function $\psi(\rho,z,t)$ to a solution of Maxwell's equation and by taking into account the spherical symmetry of the microsphere this wave function can be described in cylindrical coordinates as

$$\psi(\rho,z,t) = f(\rho,\beta)\exp\{i[(k\cos\beta)z - \omega t]\} \quad ; \quad \rho = \sqrt{x^2 + y^2} \tag{7}$$

The exponential function describes the propagation of the wave vector in the $z$ direction where $\rho$ is the distance from the symmetric z axis, $\beta$ is defined by eq. 5, and $z$ is the distance from the focal plane. We substitute eq. 7 into the wave equation

$$\nabla^2\psi(\rho,z,t) = \frac{1}{c^2}\frac{\partial^2}{\partial t^2}\psi(\rho,z,t) \quad . \tag{8}$$

Then we get [37-39]

$$\frac{d^2 f(\rho,\beta)}{d\rho^2} + \frac{1}{\rho}\frac{df(\rho,\beta)}{d\rho} + \left(k^2 - k_z^2\right) f(\rho,\beta) = 0 \tag{9}$$

We used here the azimuthal symmetry of the wave function including the relations $k^2 = \omega^2/c^2$, and $k\cos\beta = k_z$. Solution of eq. 9 is given by the Bessel function of order 0, so that

$$f(\rho,\beta) = J_0(\rho k \sin\beta) \quad ; \quad k^2 \sin^2\beta + k_z^2 = k^2 \tag{10}$$

The central bright core of the Bessel function has a radius [39]

$$\rho_0 = \frac{2.405}{k \sin\beta} \tag{11}$$

Here $\rho_0$ is the radial distance on the focal plane from the symmetric $z$ axis to the first zero of the Bessel function. As the angle $\beta$ gradually changes the central bright core radius is changed inversely proportional to $\sin\beta$. A narrow waist of the Nano-jet is obtained by the overlapping of the bright cores of the Bessel functions (with various intersecting angles $\beta$) producing in the waist a constant EM field background, so that on the waist any fluctuations by the Bessel functions are prevented. For the region beyond the waist i.e. for distances from the symmetric $z$ axis, on the focal plane, which are beyond the first zero of the Bessel functions, there will be interference between the different Bessel functions leading to oscillating EM fields. Following the present analysis I suggest to use near the focal point a circular slit with a radius $\rho_0$ which will be in the order of a subwavelength so that it will transmit the constant EM field modulated by the evanescent waves into a confocal microscope [10]. Then such system will improve the imaging of the object fine structures by the use of evanescent waves.

In the following discussion I show that the present work is in fair good agreement with measurements made on Nano-jets. Ferrand et al. [40] measured the EM field distribution in the focal plane of a dielectric sphere with micrometer dimensions illuminated by a plane wave. These measurements have shown the existence of subwavelength beam that emerges from the microsphere with high intensity and low divergence where the beam keeps its subwavelength FWHM (full width at half maximum) over a distance given by $(2-3)\lambda$. This special property: subwavelength FWHM over micron propagation distance is unreachable with a classical Gaussian laser focusing. This beam has therefore been defined as Nano-jet. Ferrand et al. [40] measured Nano-jet size as small as 270 nm FWHM for a 3 $\mu m$ sphere at a wavelength $520\,nm$. This result can be compared with the following calculation made according to the present analysis.

The EM field intensity $\Delta E$ transmitted through the microsphere for initial angles between $\theta_{in}$ and $\theta_{in}+d\theta_{in}$ is given according to eq. 4 by $F(\theta_{in})$. This field light intensity is transmitted through a circular aperture at the focal plane and it is intersecting the symmetric $z$ axis with a crossing angle $\beta$ where this angle can be calculated as function of $\theta_{in}$ by eq. 5. For the angle $\theta_{in}=45^0$ for which the transmitted field intensity $F(\theta_{in})$ is maximal, and for the indices of refraction $n_2=1.6$, $n_1 \simeq 1$ (air) given in the experiments [40], we get according to eq. (5) the result $\beta \simeq 37.54^0$. For this intersection angle and for the wavelength $520\,nm$, a straightforward calculation according to Eq. (11), for the radius of the bright central spot of the Bessel function leads to $\rho_0 \simeq 0.628\lambda \simeq 327 nm$. In making this calculation I take into account that the Nano-jet FWHM size is comparable with the radius of the bright central spot of the Bessel function. The result obtained by the present calculation is in a fair good agreement with Ferrand et al. [40], in which they obtained the waist size of 270 nm,

**A combination of the microsphere with an interferometer for phase contrast measurements**

Many objects in microscopy are phase objects which only change the phase of the incident wave without changing the amplitude. Thus, if only the refractive index or thickness of phase object varies across transverse dimensions, then by using an ordinary microscope it will not be able to observe such object. Such an object can be viewed through what is known as phase contrast microscopy [29]. Imaging of phase objects is especially important for biological systems [27, 41]. The following optical system described in Fig. 2, which is based on a combination of the microsphere with an interferometer is suggested for phase contrast measurements.

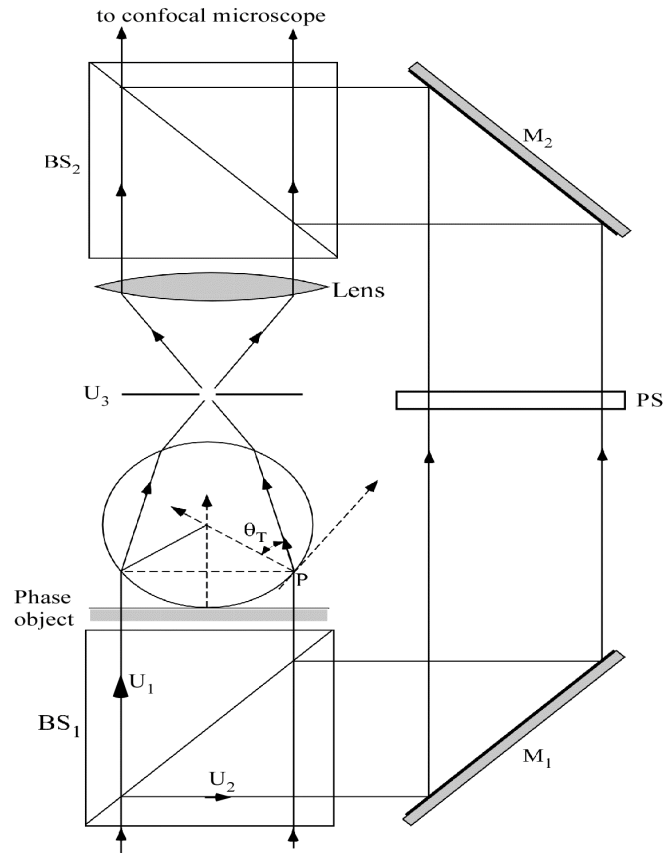

**Fig. 2**

Phase contrast measurements are described by the use of a combination of the microsphere with an interferometer. Plane EM waves are incident on the first beam splitter (BS1) and are divided into a transmitted field $U_1$, and reflected field $U_2$. The EM field $U_1$ is transmitted through a phase object, converging by the microsphere and producing a Nano-jet at the focal plane modulated by the evanescent waves, denoted in Fig. 2, as $U_3$. The EM field $U_2$ reflected from BS1 is reflected also from mirrors $M_1$ and $M_2$, and is recombined with the EM field $U_3$ at the second beam splitter (BS2). A phase shifter (PS) produces a phase difference between the two beams. The recombined beam is going into confocal microscope.

As shown in Fig. 2 plane EM wave is incident perpendicular to the first beam-splitter (BS1) where part of the light is continuing perpendicular to a thin phase object like that of a biological tissue, and a part of it is reflected into horizontal direction and reflected from mirrors M1 and M2. The EM field with constant amplitude $U_1$ after its transmittance through the thin layer of a phase object can be described as

$$U_1(x_0, y_0) = U_1 \exp\left[-i\varphi(x_0, y_0)\right] \tag{12}$$

We inserted here the phase of the object $\varphi(x_0, y_0)$ which is a function of the object $(x_0, y_0)$ coordinates. This EM field is converging by the microsphere into the focal plane and transmitted through a small circular aperture inserted in the plane barrier located on the focal plane of the microsphere.

The EM field $U_3$ on the circular aperture is composed of the Nano-jet and its modulation $\varphi(x_0, y_0)$ by the evanescent waves, as described in the previous analysis related to Fig. 1. While the EM field $U_3$ might propagate into a confocal microscope measuring fine structures light intensities, it cannot measure directly the objects phases. For the purpose of measuring phase fine structures we use the second EM field $U_2$ reflected from BS1 that after reflection from mirrors M1 and M2 and transmittance through a phase shifter (PS) is given by $U_2 \exp(-i\alpha)$. By using the phase shifter (PS) the phase difference between the beams $U_2 \exp(-i\alpha)$ and $U_3$ can be changed. The EM fields $U_3$ and $U_2 \exp(-i\alpha)$ are recombined by the second beam splitter (BS2) leading to interference between the two beams with light intensity $|U_4|^2$ which can be described schematically as

$$|U_4(x_4, y_4)|^2 = |CU_3|^2 + |U_2|^2 + 2|CU_3||U_2|\cos\left[\varphi(Mx_0, My_0) + \alpha\right] \tag{13}$$

Here the constant C represents the relative intensity between the two beams. $\varphi(Mx_0, My_0)$, represents the phases of the phase object where in a simple geometric approach they can be magnified by factor M. $\alpha$ represents the phase difference between the two beams which is controlled by the phase shifter PS. The conversion of the phases of the phase object to light intensities is demonstrated by the physical scheme described Fig. 2 and the schematic analytical relation given by eq. 13. The realization of such phase measurements is especially important for biological systems but so far such proposed system has not been exploited.

**Conclusions**

The Nano-jet in microsphere optical experiments is found to be produced by propagating Bessel beams. By using a plane barrier in the focal plane with a circular aperture with a radius in the order of subwavelength dimensions the oscillations produced in the Nano-jet are eliminated, as all the bright cores of the various Bessel beam are overlapping. The information in this system is obtained from modulation of the Nano-jets by evanescent waves transmitted through the microsphere. This system is expected to give very high resolutions which might be applied in various fields of integrated optics.

A combination of the microsphere with an interferometer is described for phase contrast measurements, as analyzed in the present work. Such system might be used for imaging phase objects and is especially important for imaging biological systems.

**Acknowledgement**

The author would like to thank Prof. S. Lipson for interesting discussions.

**Authors' contributions**

The author approves the final original paper written by him.

**Competing interests**

The author declares he has no competing interests.